\documentclass[twocolumn]{aastex63}

\usepackage{float}
\usepackage{gensymb}

\shorttitle{Evolution of Li in red clump stars}
\shortauthors{$\rm Singh\ et\ al.\ 2021$}

\begin{document}

\title{Tracking the evolution of lithium in giants using asteroseismology: Super-Li-rich stars are almost exclusively young red-clump stars}

\correspondingauthor{Raghubar Singh}
\email{raghubar2015@gmail.com}

\author{Raghubar Singh}
\affil{Indian Institute of Astrophysics, 560034, 100ft road Koramangala, Bangalore, India}
\affil{Pondicherry University R. V. Nagara, Kala Pet, 605014, Puducherry, India}

\author{Bacham E. Reddy}
\affil{Indian Institute of Astrophysics, 560034, 100ft road Koramangala, Bangalore, India}

\author{Simon W. Campbell}
\affil{School of Physics and Astronomy, Monash University, Clayton, Victoria, Australia}

\author{Yerra Bharat Kumar}
\altaffiliation{LAMOST Fellow}
\affil{Key Laboratory for Optical Astronomy, National Astronomical Observatories, Chinese Academy of Sciences, Beijing, 100101, China} 

\author{Mathieu Vrard} 
\affil{Dept.  of  Astronomy,  The  Ohio  State  University,  140  W.  18th  Ave., Columbus, OH 43210, USA} 
\begin{abstract}
We report novel observational evidence on the evolutionary status of lithium-rich giant stars by combining asteroseismic and lithium abundance data. Comparing observations and models of the asteroseismic gravity-mode period spacing $\Delta\Pi_{1}$, we find that super-Li-rich giants (SLR, A(Li)~$> 3.2$~dex) are almost exclusively young red-clump (RC) stars. Depending on the exact phase of evolution, which requires more data to refine, SLR stars are either (i) less than $\sim 2$~Myr or (ii) less than $\sim40$~Myr past the main core helium flash (CHeF). Our observations set a strong upper limit for the time of the inferred Li-enrichment phase of $< 40$~Myr post-CHeF, lending support to the idea that lithium is produced around the time of the CHeF. In contrast, the more evolved RC stars ($> 40$~Myr post-CHeF) generally have low lithium abundances (A(Li)~$<1.0$~dex). Between the young, super-Li-rich phase, and the mostly old, Li-poor RC phase, there is an average reduction of lithium by about 3 orders of magnitude. This Li-destruction may occur rapidly. We find the situation to be less clear with stars having Li abundances between the two extremes of super-Li-rich and Li-poor. This group, the `Li-rich' stars ($3.2 >$~A(Li)~$> 1.0$~dex), shows a wide range of evolutionary states. 

\end{abstract}

\keywords{ stars: abundances --- stars: evolution --- stars: interiors --- stars: low-mass}

\section{Introduction}

The origin of strong lithium abundance excess in red giants - `Li-rich giants' - has been a long-standing problem ever since its discovery about four decades ago \citep{wallerstein1982}. In the last few years significant progress has been made thanks primarily to large datasets of ground-based spectroscopic  surveys such as LAMOST \citep{Cui2012} and GALAH \citep{galah2015}, and the space-based {\it Gaia} astrometric \citep{Gaiadr12016, Gaiadr22018} and Kepler time-resolved photometric surveys \citep{Borucki2010}. Studies using data from these surveys have now confirmed the early suspicions (eg. \citealt{Kumar2011, SilvaAguirre2014}) that Li-rich giants are predominantly in the He-core burning phase of stellar evolution, also known as the red clump (RC; \citealt{ Singh2019l, Singh2018l, Smiljanic2018, deepak2019, Casey2019}). This new development has significantly narrowed the search for finding the origin of the lithium enhancement. 

It is also well established that Li is almost totally destroyed during the phase of evolution just preceding the RC, the red giant branch (RGB; \citealt{Lind2009a, Bharatn2020}), reaching values as low as A(Li)~$\sim -1.0$~dex. In contrast, the average Li abundance on the RC is A(Li)~$= +0.7$~dex \citep{Bharatn2020}, implying  that there must be a lithium production phase between the late RGB and RC. Since that study two theoretical models have been put forward for Li production. The first is by \cite{Mori2020} whose model produces Li at the RGB tip, before the onset of He-flash, via the inclusion of the neutrino magnetic moment. The second proposed model produces Li during the main CHeF, by assuming some ad-hoc mixing during the extremely energetic flash \citep{Schwab2020}. Both the models explain the observations, with average RC A(Li) of $+0.7$~dex, however they do not attempt to explain the (super)Li-rich giants. That said, variation in the parameters of the \citealt{Schwab2020} model can indeed produce the very high Li abundances required (see their Fig.~4), although the author notes that these high Li abundances are quickly depleted in their model. The current study attempts to better isolate the location of the Li production site as stars transition from the RGB tip to RC phase. We also aim to explore Li evolution along the RC.

We do this by using observations of asteroseismic parameter $\Delta\Pi_{1}$, the asymptotic gravity-mode period spacing of dipole modes \citep{Unno1989,Mosser2012cat,Mosser2014, Vrard2016}, which has been shown to clearly vary with evolution in stellar models of the RC phase \citep{Bildsten2012, Constantino2015, Bossini2015}. We combine $\Delta\Pi_{1}$ with newly derived Li abundances as well as Li abundances from the literature.

\section{Sample selection}
\label{sec:sample}

We extracted a sample of 6955 low mass (M~$\le 2$~M$_{\odot}$) giants that are classified as RC stars based on Kepler asteroseismic data \citep{Yu2018}. We then searched the literature for $\Delta\Pi_{1}$ values \citep{Mosser2012cat, Mosser2014, Vrard2016} and Li abundances \citep{Tajitsu2017, Singh2018l, Singh2019l, Yan2020n}. We found 37 stars with A(Li) and $\Delta\Pi_{1}$, and we measured $\Delta\Pi_{1}$ in 10 more (Sec.~\ref{sec:dpi1}), giving a total of 47 RC stars with both parameters.

We also searched for Kepler field stars in the recently released catalogue of medium resolution (MRS, R~$\approx$~7,500) spectroscopic data of the LAMOST\footnote{\url{http://dr6.lamost.org/}} survey \citep{lamostmedres2020}, finding 577 RC stars. We were able to measure Li in 22 of these stars (Sec.~\ref{sec:mrsabunds}). Of these, we found $\Delta\Pi_{1}$ for 6 of them in the literature \citep{Mosser2012cat, Mosser2014, Vrard2016} and were able to measure $\Delta\Pi_{1}$ for a further 6 of them (Sec.~\ref{sec:dpi1}), giving an extra 12 stars with both parameters.

\begin{figure}
\includegraphics[width=0.47\textwidth]{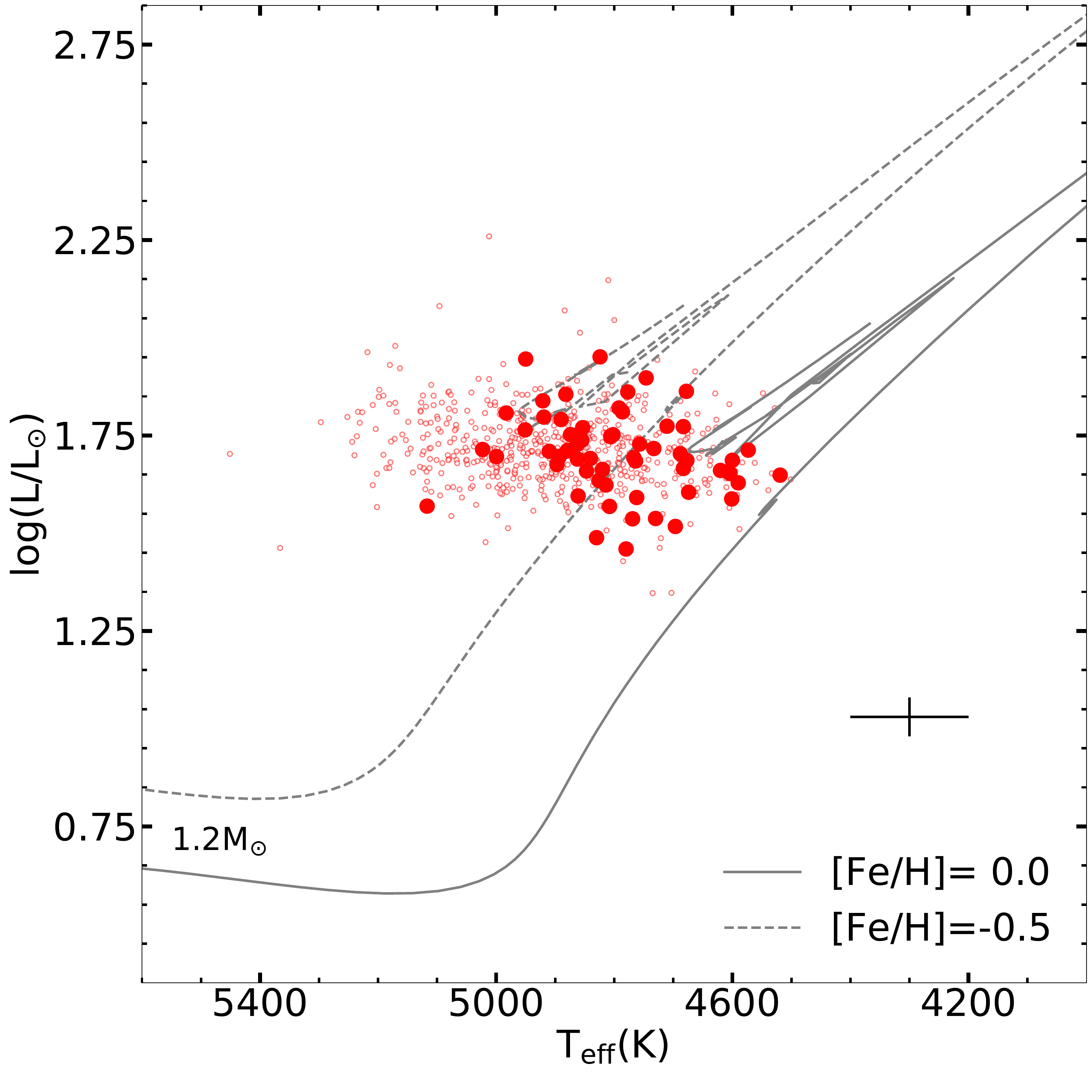}
\caption{HR diagram showing the sample of 59 RC giants for which we have $\Delta\Pi_{1}$ and A(Li) (large symbols). For context we show the entire Kepler-LAMOST MRS overlap RC sample of 577 giants in the background. Overplotted are two stellar model tracks, both 1.2~M$_{\odot}$ but with two different metallicities. Typical observational uncertainties are indicated by the error cross.}
\label{fig:lumteff}
\end{figure}

Combining this sample with the literature sample, we have a total of 59 RC stars with certain evolutionary phase and accurate Li abundances, all from the Kepler field. Luminosities were derived using the {\it Gaia} distances \citep{Gaiadr22018} and the bolometric correction from \cite{Torres2010}. We discuss possible biases in our sample in Appendix~\ref{appx:bias}. We found no bias in our sample that could alter our conclusions. The final sample is shown in Figure~\ref{fig:lumteff}.

\section {Results}
\label{sec:results}

\subsection{LAMOST MRS Lithium abundances}
\label{sec:mrsabunds} 

\begin{figure}[t]
\includegraphics[width=0.46\textwidth]{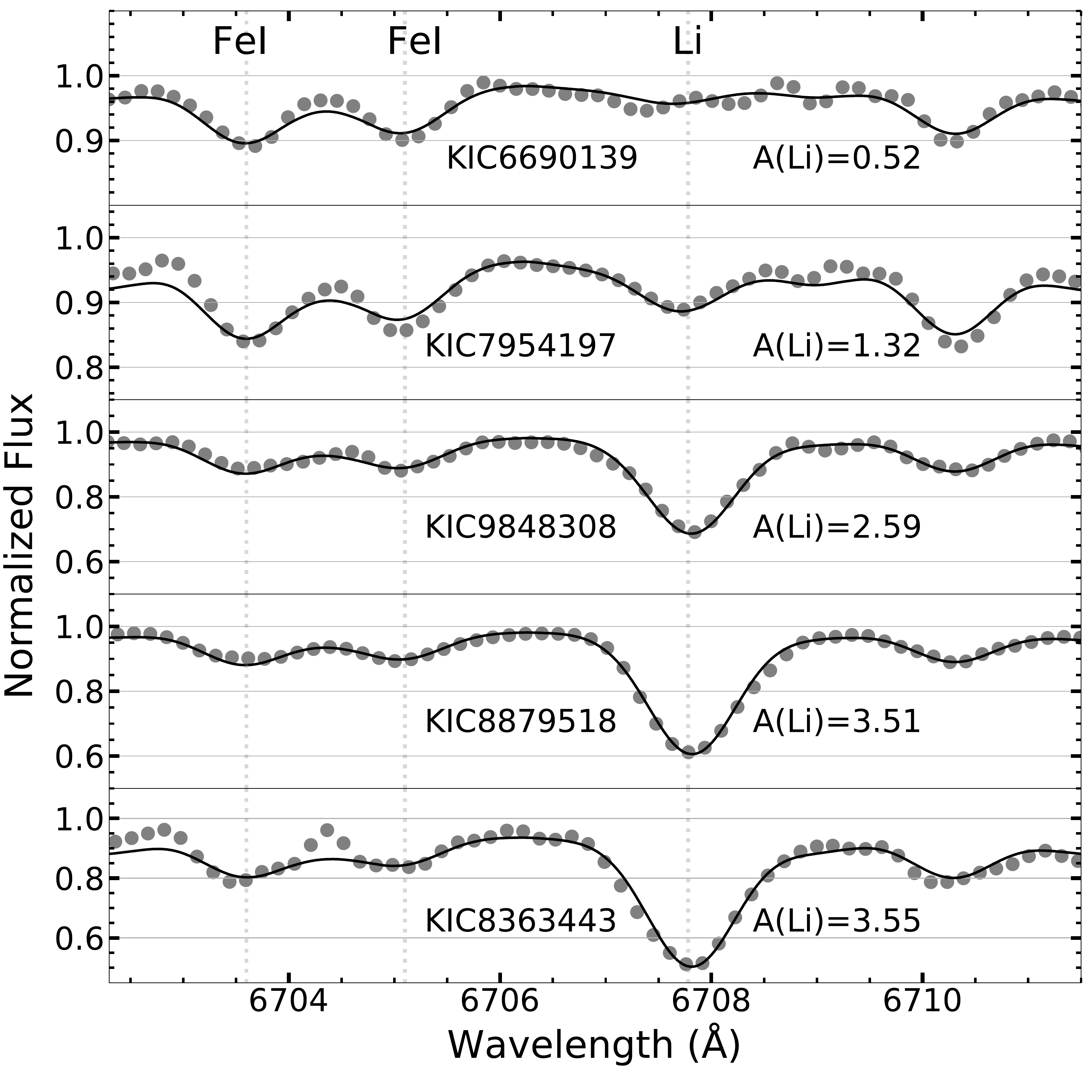}
\caption{Derivation of Li abundances for a few representative giants using spectral synthesis of the 6707.78~\AA\ Li line (solid lines) in the LAMOST medium resolution spectra (dots). 
}
\label{fig:lisynth}
\end{figure}

Li abundances were derived by matching the synthetic spectra with the the observed Li line at 6707.78~\AA~in the medium-resolution LAMOST spectra.  We used the 2013 version of the local thermal equilibrium (LTE) radiative transfer code MOOG \citep{Sneden1973}, in combination with the Kurucz ATLAS atmospheric models \citep{Castelli2003}. Atmospheric parameters (T$_{\rm eff}$, log~g, [Fe/H]) were taken from the \cite{Yu2018} catalogue. Our line list, with atomic and molecular data, was compiled using the Linemake code\footnote{\url{https://github.com/vmplacco/linemake}}. Figure~\ref{fig:lisynth} shows a comparison of synthetic spectra with the observed spectra of a few representative stars. The estimated uncertainty in Li abundance is the quadratic sum of uncertainties in Li abundance caused by the estimated uncertainties in atmospheric parameters for each individual star (see Table~\ref{tab:tab1}).  Li abundances were corrected for non-LTE) effects using the $\Delta_{\rm{NLTE}}$ values provided by \cite{Lind2009a}. 

We estimated the lower detectability limit in equivalent width (EW) for our data using the \cite{Cayrel1988} formulation for uncertainties in EW for a range of SNR = 55 -- 150. To err on the conservative side, we adopted a lower limit of $3\times$ $\delta \rm EW$ = (10.6 -- 28.2)~m\AA, which translates to lower limit abundances in the range +0.7 to +1.1 dex, depending on stellar parameters (Fig.~\ref{fig:lisynth}). This criterion resulted in 22 RC stars for which we could reliably derive Li abundances, for 12 of which we have $\Delta\Pi_{1}$ (Table~\ref{tab:tab1}).

\begin{table*}
\centering
\begin{tabular}{ccccccclr}
\hline
KIC &  V\footnote{\url{https://simbad.u-strasbg.fr/simbad/sim-fbasic}} & T$_{\rm eff}$ & log~g & [Fe/H]& log(L/L$_{\odot}$) & A(Li)$_{\rm NLTE}$& M(M$_{\odot}$) & $\Delta \Pi_{1}$   \\
\hline
\multicolumn{9}{c}{Sample from this work of LAMOST-MRS and Kepler Field} \\
\hline
5079905 &12.56 & 4769$\pm$139 &2.38$\pm$0.01 & 0.03$\pm$0.30 &1.54$\pm$0.04 &2.66$\pm$0.21$^{1}$&0.92$\pm$0.09   & 298$\pm$20$^{1}$   \\ 
6593240 &11.11 & 4808$\pm$80  &2.41$\pm$0.01 & 0.0$\pm$0.15  &1.57$\pm$0.03 &2.74$\pm$0.15$^{1}$&0.83$\pm$0.09   & 277$\pm$20$^{2}$    \\
7020392 &12.40 & 4893$\pm$100 &2.39$\pm$0.01 &-0.25$\pm$0.15 &1.83$\pm$0.04 &1.45$\pm$0.16$^{1}$&0.97$\pm$0.12   & 299$\pm$20$^{1}$    \\
7595155 &12.12 & 4601$\pm$100 &2.36$\pm$0.01 &0.30$\pm$0.30  &1.74$\pm$0.03 &1.23$\pm$0.20$^{1}$&0.86$\pm$0.08   & 291$\pm$17$^{2}$    \\
7612438 &12.88 & 4921$\pm$80  &2.54$\pm$0.01 &0.07$\pm$0.15  &1.96$\pm$0.08 &1.86$\pm$0.15$^{1}$&1.87$\pm$0.10   & 279$\pm$32$^{1}$    \\
7954197 &10.90 & 4814$\pm$139 &2.42$\pm$0.01 &0.12$\pm$0.30  &1.62$\pm$0.03 &1.32$\pm$0.20$^{1}$&1.23$\pm$0.12   & 269$\pm$2$^{1}$    \\  
8363443 &10.88 & 4606$\pm$80  &2.41$\pm$0.01 &0.38$\pm$0.15  &1.81$\pm$0.03 &3.55$\pm$0.12$^{1}$&1.33$\pm$0.09   & 248$\pm$2$^{1}$      \\
8619916 &12.90 & 5117$\pm$160 &2.44$\pm$0.01 &-0.05$\pm$0.30 &1.71$\pm$0.06 &2.71$\pm$0.25$^{1}$&1.11$\pm$0.11   & 300$\pm$3$^{2}$     \\
8879518 &11.22 & 4863$\pm$80  &2.57$\pm$0.01 &0.14$\pm$0.15  &1.73$\pm$0.03 &3.51$\pm$0.12$^{1}$&1.76$\pm$0.13   & 268$\pm$3$^{2}$     \\ 
9848308 &12.28 & 4733$\pm$80  &2.49$\pm$0.01 &0.06$\pm$0.15  &1.72$\pm$0.04 &2.59$\pm$0.12$^{1}$&1.36$\pm$0.09   & 329$\pm$4$^{2}$     \\
9907856 &12.45 & 4855$\pm$80  &2.41$\pm$0.01 &-0.17$\pm$0.15 &1.73$\pm$0.01 &2.56$\pm$0.13$^{1}$&1.16$\pm$0.25   & 328$\pm$3$^{2}$     \\ 
11129153&12.09 & 4830$\pm$100 &2.37$\pm$0.03 &-0.37$\pm$0.15 &1.63$\pm$0.08 &1.45$\pm$0.16$^{1}$&1.12$\pm$0.21   & 228$\pm$11$^{1}$     \\  
\hline
\multicolumn{9}{c}{Sample from literature for which $\Delta \Pi_{1}$ values are derived in this study} \\
\hline
3751167 &13.74 & 4914$\pm$80  &2.33$\pm$0.03 &-0.76$\pm$0.15 &1.86$\pm$0.06 &4.0$\pm$0.35$^{3}$ & 0.95$\pm$0.22  & 260.0$\pm$13$^{1}$\\
7131376 &13.99 & 4833$\pm$152 &2.45$\pm$0.01 & 0.04$\pm$0.30 &1.52$\pm$0.06 &3.80$\pm$0.35$^{3}$& 1.25$\pm$0.11  & 250.8$\pm$18$^{1}$\\
7899597 &13.61 & 4757$\pm$80  &2.40$\pm$0.03 &-0.10$\pm$0.15 &1.77$\pm$0.05 &3.39$\pm$0.06$^{4}$& 1.28$\pm$0.23  & 267.4$\pm$18$^{1}$\\
8869656 &9.34  & 4915$\pm$137 &2.40$\pm$0.01 &-0.13$\pm$0.30 &1.68$\pm$0.02 &3.61$\pm$0.09$^{4}$& 1.21$\pm$0.13  & 235.5$\pm$13$^{1}$\\
9667064 &13.35 & 4802$\pm$80  &2.38$\pm$0.03 & 0.04$\pm$0.15 &1.86$\pm$0.06 &4.40$\pm$0.35$^{3}$& 1.40$\pm$0.20  & 260.9$\pm$16$^{1}$\\
9833651 &12.52 & 4730$\pm$80  &2.49$\pm$0.01 & 0.14$\pm$0.15 &1.66$\pm$0.04 &3.40$\pm$0.09$^{4}$& 1.50$\pm$0.13  & 266.4$\pm$3$^{1}$\\
11615224&11.15 & 4888$\pm$100 &2.38$\pm$0.01 &-0.03$\pm$0.15 &1.89$\pm$0.05 &2.84$\pm$0.02$^{4}$& 0.85$\pm$0.10  & 257.7$\pm$2$^{1}$\\
11658789&13.35 & 5226$\pm$155 &2.42$\pm$0.02 &-0.52$\pm$0.30 &1.69$\pm$0.03 &3.90$\pm$0.35$^{3}$& 0.87$\pm$0.14  & 256.0$\pm$2$^{1}$\\
11805390 &9.78 & 4950$\pm$80  &2.58$\pm$0.01 &-0.06$\pm$0.15 &1.65$\pm$0.03 &2.24$\pm$0.10$^{4}$& 1.34$\pm$0.10  & 247.9$\pm$2$^{1}$\\
12784683&11.46 & 4983$\pm$145 &2.37$\pm$0.02 &-0.30$\pm$0.30 &1.68$\pm$0.02 &2.79$\pm$0.08$^{4}$& 1.16$\pm$0.18  & 282.3$\pm$18$^{1}$\\
\hline
\end{tabular}
\caption{ RC sample of 22 stars (out of total 59) for which A(Li) and/or $\Delta \Pi$  are measured in this work (see for details Section 2). Masses were derived using the relation given in \cite{Bedding1995}.\\ \\
Notes: $^{1}$ This work; $^{2}$ \cite{Vrard2016}; $^{3}$ \cite{Singh2019l}; $^{4}$  \cite{Yan2020n} 
}
\label{tab:tab1}
\end{table*}

\subsection{Asymptotic gravity-mode period spacing $\Delta\Pi_{1}$}
\label{sec:dpi1}

Of the 59 RC giants in our sample, $\Delta \Pi_{1}$ values for 43 are available in the literature \citep{Mosser2012cat,Mosser2014,Vrard2016}. For the other 16 we derived it ourselves. We downloaded the time series data from the Kepler archive and converted them to the frequency domain using the Lightkurve package\footnote{\url{https://github.com/NASA/Lightkurve}} \citep{lightkurve2018}. Details of our method of $\Delta\Pi_{1}$ estimation are given in \cite{Mosser2014} and \cite{Vrard2016}. We were able to derive asymptotic values  (Table~\ref{tab:tab1}) for 16 RC giants which have have sufficient SNR in their power density spectra.

\subsection{A(Li) vs $\Delta\Pi_{1}$: Evolution of Li along the RC}

\begin{figure}[!t]
\includegraphics[width=0.45\textwidth]{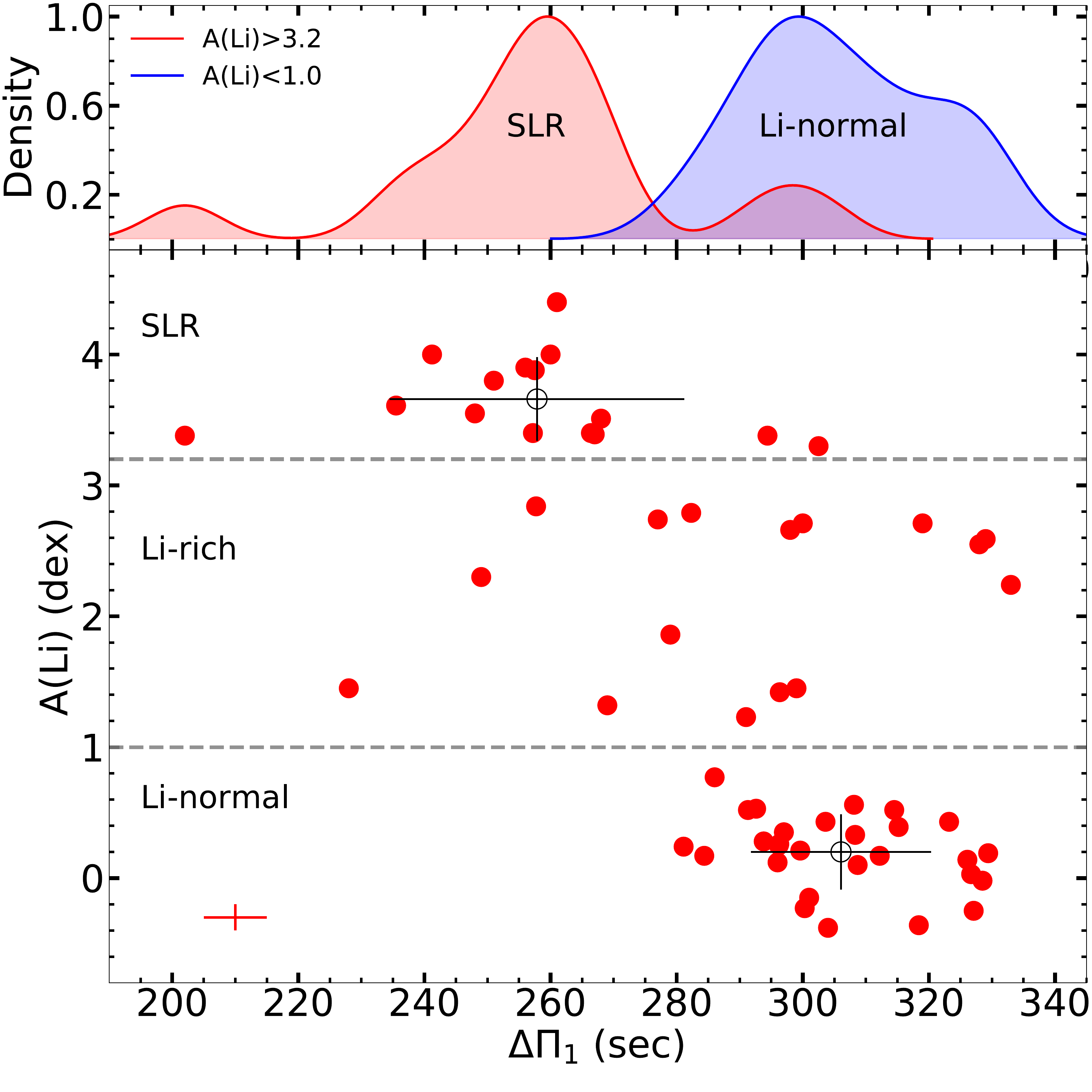}
\caption{Variation of A(Li) with asymptotic gravity mode period spacing $\Delta\Pi_{1}$ in RC stars. The open circles with error bars show the mean and $1\sigma$ dispersion of the groups. Due to the broadness of the Li-rich group we do not assign a mean to it. The top panel shows Gaussian kernel density histograms for the SLR and LN groups. Since $\Delta\Pi_{1}$ tracks the evolution of the stars (from left to right, see Fig.~\ref{fig:model1}), it can be seen that the average Li abundance reduces with evolution. Typical A(Li) and $\Delta\Pi_{1}$ uncertainties are represented by the error cross at bottom left.}
\label{fig:alipg}
\end{figure}

In Figure~\ref{fig:alipg} we display A(Li) vs $\Delta\Pi_{1}$ for the RC stars. We divide the sample into three groups depending on their Li abundance: 

\begin{itemize}
\item Li-normal (LN): A(Li)~$< 1.0$~dex
\item Li-rich (LR): $1.0 < $~A(Li)~$< 3.2$~dex
\item Super-Li-rich (SLR): A(Li)~$> 3.2$~dex
\end{itemize}

The A(Li)~$= 1.0$~dex delineation is based on the \cite{Bharatn2020} study's distribution on the RC (peaked at $ +0.7$~dex), whilst the A(Li)~$ = 3.2$~dex delineation is based on the ISM value \citep{Knauth2003}. Our key result emerges from Figure~\ref{fig:alipg} -- the great majority ($12/15 = 80\%$) of the SLR stars have $\Delta\Pi_{1}$ peaked at low values (mean~$= 257 \pm 23$~s). We note that two of the three outliers in the SLR group have masses $2\sigma$ from the mean mass of our RC sample ($\sim 1.8~\rm{M}_{\odot}$, where the sample mean mass is $1.2\pm 0.3~\rm{M}_{\odot}$). Without these three outliers\footnote{We do not remove these stars from the sample, this is only a test of the sensitivity of the dispersion, since with the outliers the distribution is non-Gaussian.} the average remains the same, at 255~s, but the dispersion is halved ($\sigma = 10$~s), as expected from the histogram. In contrast, the Li-normal stars all have higher $\Delta\Pi_{1}$, ranging from 280~s up to $\sim 330$~s. Their mean is $306\pm 14$~s, much higher than the SLR group.

The Li-rich stars (Fig.~\ref{fig:alipg}) show more complex behaviour. They have a very broad distribution of $\Delta\Pi_{1}$ values rather than forming a coherent group. We now discuss the implications of the $\Delta\Pi_{1}$-A(Li) observations in relation to stellar evolution.

\begin{figure}[t]
\includegraphics[width=\columnwidth]{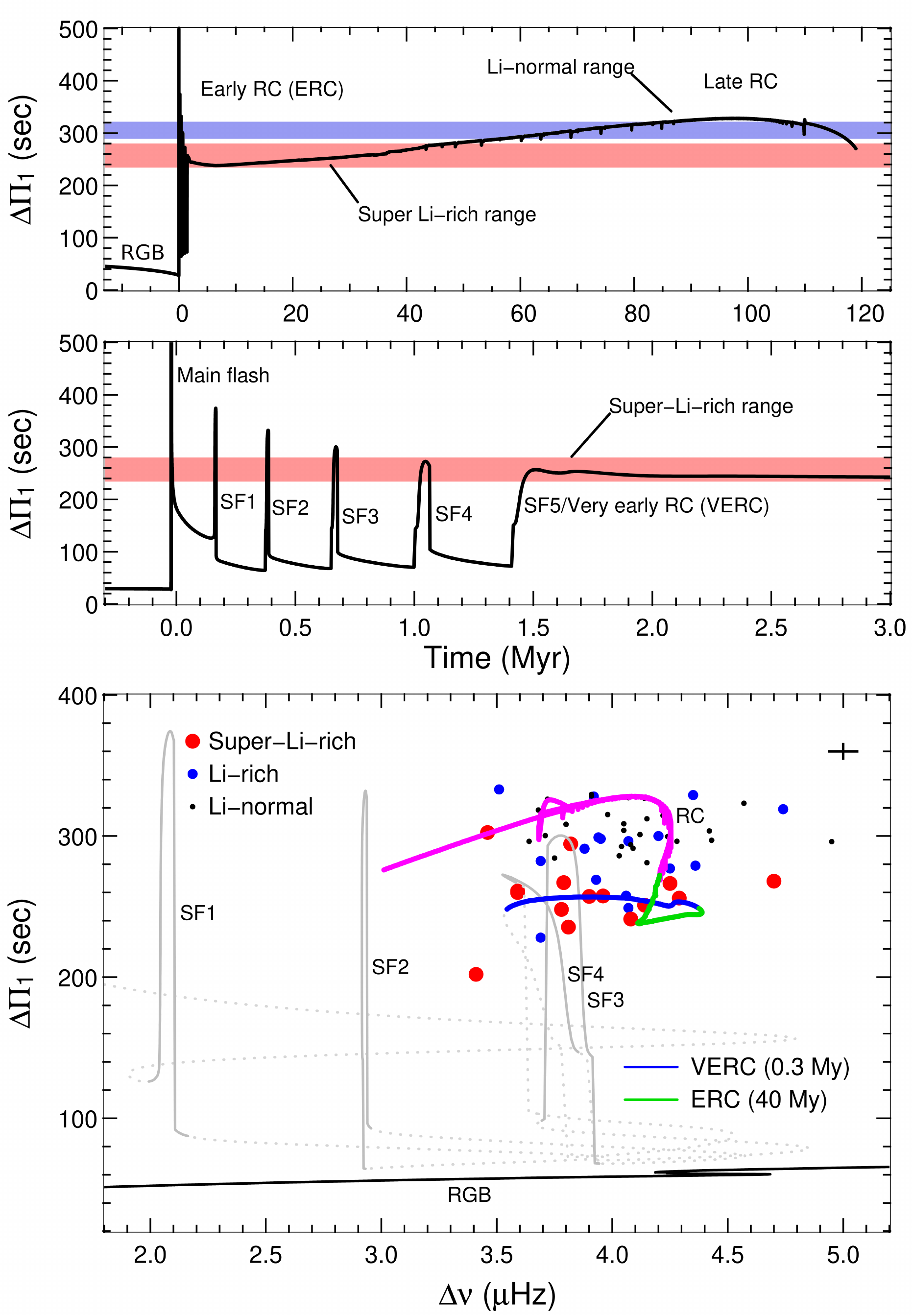}
\caption{Stellar model showing evolution of  $\Delta\Pi_{1}$ and $\Delta\nu$. {\bf Top panel:} $\Delta\Pi_{1}$ evolution through the late RGB, the brief helium core flash phase, and the entire RC phase. The shaded bands show the $1\sigma$ intervals of $\Delta\Pi_{1}$ for the SLR (red) and Li-normal stars (blue), corresponding to the distributions in Fig.~\ref{fig:alipg}. {\bf Middle panel:} Same as above but magnified to focus on the helium flashing phase and the early RC. {\bf Bottom panel:} Model evolution in the $\Delta\Pi_{1}$-$\Delta\nu$ plane (lines), compared to observations. Dotted lines are used for the evolution between (sub)flashes for clarity. }
\label{fig:model1}
\end{figure}

\section{Discussion} 
\label{sec:discussion}

\subsection{Observations versus Stellar Evolution}

Low mass giants ($\leq$ 2~M$_{\odot}$) develop electron-degenerate cores while ascending the red giant branch. Initiation of an off-centre helium flash in the degenerate core terminates the RGB evolution \citep{Schwarzchild1962,Demarque1971,sweigart1994}. The first/main He-flash is very short-lived, lasting about 1000 years. After the main flash a series of progressively weaker flashes ensues (Fig.~\ref{fig:model1}; also see \citealt{thomas1967, Cassisi2003,Bildsten2012,Deheuvels2018}), each igniting closer and closer to the centre, progressively removing the electron degeneracy in the core. The entire He-flashing phase lasts $\sim 1.5-2$~Myr. Finally the very centre of the core becomes convective, marking the start of the RC phase (quiescent core helium burning), which lasts for $\sim 100$~Myr. Throughout this evolution $\Delta\Pi_{1}$ varies predictably, as seen in Fig.~\ref{fig:model1} (also see \citealt{Bildsten2012,Constantino2015,Bossini2015,Deheuvels2018}). Our new observational data enables us to compare the observed $\Delta\Pi_{1}$ with model predictions.

We have calculated a series of stellar models using the MESA code (\citealt{paxton2011,paxton2019}; version 12778) which vary in mass\footnote{We note that the seismic masses are the current masses of the stars. The total RGB mass loss is $\sim 0.1$~M$_{\odot}$, which is small compared to the dispersion of our sample ($0.3$~M$_{\odot}$).} (M~$= 0.9, 1.2, 1.4$~M$_{\odot}$) and metallicity ([Fe/H]$= -0.4, +0.0$), to cover the bulk of our observational sample. RGB mass loss was modelled using the \cite{reimers1975} formula ($\eta = 0.3$). We used the standard MESA nuclear network (`basic.net') and standard equation of state (see \citealt{paxton2019} for details), and $\alpha_{MLT} = 2.0$. Convective boundary locations were based on the Schwarzschild criterion, extended with exponential overshoot (\citealt{herwig1997}). 
In Figure~\ref{fig:model1} we show a representative model that matches the mean mass of our sample, 1.2~M$_{\odot}$, and is of solar metallicity\footnote{Solar abundances were from \cite{asplund2019}} (the mean [Fe/H] of our sample is $-0.14\pm 0.26$~dex). As \cite{bossini2017} showed, the very early RC $\Delta\Pi_{1}$ evolution is not dependent on mass, however it has some dependence on metallicity. Comparing the dispersion of [Fe/H] in our sample (0.26~dex) with the models in Figure~3 of \cite{bossini2017}, we expect a variation of $\sim 5$~s in $\Delta\Pi_{1}$. This is approximately the same as our observational error bars (Fig.~\ref{fig:alipg}). On the other hand, the late RC is particularly sensitive to the treatment of convective boundaries (eg. \citealt{Constantino2015,Bossini2015}). To match the highest $\Delta\Pi_{1}$ values ($\sim 330$~s) we adopted an overshoot parameter $f_{OS} = 0.02$.

We display the model evolution of $\Delta\Pi_{1}$ versus time in the top panels of Figure~\ref{fig:model1}. Overplotted are the $1\sigma$ intervals of $\Delta\Pi_{1}$ for the observations of the SLR and Li-normal groups (see Fig.~\ref{fig:alipg}). It can be seen that the high values of $\Delta\Pi_{1}$ of the Li-normal stars are only available in the second half of the RC evolution (from $\sim 40$~Myr onwards), or for extremely brief periods ($\sim 0.005$~Myr) during the early helium flashes, which are unlikely to be observed in such numbers. We thus identify the Li-normal stars as more evolved RC stars.

In contrast, the low $\Delta\Pi_{1}$ values of the SLR stars are only consistent with young RC stars ($\lesssim 40$~Myr in the 1.2~M$_{\odot}$ model). Interestingly we find there is very little overlap between the Li-normal and SLR populations. This may indicate a rapid Li-depletion event (but see Sec.~\ref{subsec:linormal}).

\subsection{Are SLR Stars He-flash Stars?}
\label{subsec:heflash}

One of the main hypotheses for the Li-enrichment event is that it occurs during the main helium flash (\citealt{Mocak2011,Kumar2011,SilvaAguirre2014,Bharatn2020,Schwab2020}). Here we explore the possibility that our sample of SLR stars are currently experiencing the CHeF, or are in the following subflashing phases.

In Figure~\ref{fig:model1} it can be seen that there is some degeneracy in $\Delta\Pi_{1}$ during the early phases of the RC evolution. This makes identifying the exact young RC phase that the SLR stars reside in more difficult.

The vertical portions of the $\Delta\Pi_{1}$ evolution are extremely short-lived, so it is very unlikely to find stars in these phases. Adding the extra dimension of $\Delta\nu$ (bottom panel of Fig.~\ref{fig:model1}), we see that the $\Delta\nu$ values of the model's early subflashes are inconsistent with our sample. This is also true of the main CHeF (not pictured), which has $\Delta\nu \sim 0.04~\mu$Hz. In addition, the main flash has $\Delta\Pi_{1} = 700$~s, well outside our observed range. Thus we can rule out our SLR stars being in the main core helium flash. The third subflash (SF3) has $\Delta\Pi_{1} \sim 300$~s, well above the SLR average value of 257~s, so we rule this out also. We checked the variation of $\Delta\Pi_{1}$ during the subflashes in our small set of models and found only small deviations of $\sim \pm 5$~s in our mass and metallicity range\footnote{The number of SFs remain the same (five), but there are some small differences in timescales, dependent on mass and metallicity. The similarity is due to the He cores all being quite similar, which also gives rise to the RC itself}. This variation is similar to our uncertainties on $\Delta\Pi_{1}$. Three possibilities remain, defined by the SLR $\Delta\Pi_{1}$ band:

\begin{itemize}
\item The first $\sim 40$~Myr of RC evolution, `early RC' (ERC in Fig.~\ref{fig:model1});
\item The first $\sim 0.3$~Myr of evolution, which encompasses the `merged' subflash SF5, named `very early RC' (VERC: $1.6 -1.9$~Myr range in Fig.\ref{fig:model1});
\item The final separated helium subflash, SF4, which has a lifetime of 0.04~Myr and $\Delta\nu$ values within the range of our observations.
\end{itemize}

Large surveys give us statistical information on Li-rich giants that may also help in identifying the exact phase of SLR stars. Using the large samples of \cite{Bharatn2020} and \cite{Singh2019l} we have calculated the SLR fractions to be $0.3\%$ and $0.5\%$, respectively. If we hypothesise that all stars go through a SLR phase, then we can estimate timescales of the SLR phase. The RC phase lasts for $\sim100$~Myr, implying that the SLR phase would last around $0.3-0.5$~Myr. The lifetime of SF4 is only 0.04~Myr, so this is inconsistent with the timescale estimated from the surveys. Put another way, we find too many SLR stars for the SF4 scenario, by a factor of $\sim 10$. Further, we estimated the probability of finding even one SF star in our sample and found an expectation of 0.02 stars. Given these various lines of evidence we conclude that our SLR stars are highly unlikely to be in a flashing phase.

\subsection{How young are the SLR stars?}
\label{subsec:heflash}

To attempt to distinguish between the two remaining phases we turn to our series of models, to quantify the theoretical versus observational dispersion in $\Delta\Pi_{1}$ and $\Delta\nu$. We found that the various stellar tracks covered the whole observed wide range of $\Delta\nu$ for the SLR stars. Other uncertain model parameters, such as the convective mixing length $\alpha_{MLT}$, would further increase the dispersion in $\Delta\nu$ (\citealt{Constantino2015,bossini2017}). Thus we are unable to distinguish between the VERC and ERC scenarios using $\Delta\nu$.

In contrast, as mentioned above, $\Delta\Pi_{1}$ does not vary much with mass in the early RC, as reported by \cite{bossini2017}. During the VERC phase (not reported by \citealt{bossini2017}) we find a small dispersion in our set of models. The variation with mass is $\sim \pm 5$~s, and with metallicity $\sim \pm 10$~s, which is comparable to the observed dispersion of the SLR stars ($\sigma = 23$~s; or $\sigma = 10$~s if outliers are ignored, see Sec.~\ref{sec:results}). Thus, within the mass and metallicity variation of our sample, the ERC and VERC phases always present low $\Delta\Pi_{1}$ values, close to what we observe in the SLR stars. This reinforces the finding that these objects are all young, or very young, RC stars.

Unfortunately, due to the $\Delta\nu$ and $\Delta\Pi_{1}$ degeneracy we can not distinguish between these two phases. We now briefly discuss the implications of both scenarios.

In the ERC case only a small proportion, $\sim 1\%$, of stars would be super-Li-rich. This is a factor of 2-3 higher than the fraction found when taking into account all RC stars, consistent with the SLR stars being only found in the first $\sim40\%$ of the RC lifetime. In the VERC case $\sim 100\%$ of low-mass stars would be SLR. That is, super-Li-richness would be a universal phase of low-mass stars. They would already be super-Li-rich as they start the RC, which would be consistent with the model of \cite{Schwab2020}, where the Li-enrichment occurs during the main CHeF. A universal SLR phase would also resonate with the finding of \cite{Bharatn2020}, who report that all low-mass stars appear to go through a Li-enrichment phase, albeit to more moderate abundances -- although this difference would be explained by our Li-normal group, which indicates that strong depletion occurs during the RC.

\subsection{The Li-rich group}
\label{subsec:linormal}

The Li-rich group (Fig.~\ref{fig:alipg}) is more difficult to interpret, since the $\Delta\Pi_{1}$ distribution is so broad. We speculate that these stars could be in a phase of evolution intermediate to the SLR and Li-normal stars, currently depleting Li on their way to the late RC. This aligns with their $\Delta\Pi_{1}$ distribution being peaked at 290~s, which is between that of the SLR group (257~s) and the Li-normal group (306~s). If the Li-rich stars are currently undergoing Li-depletion then it suggests the Li depletion is a slower process. More information is needed to disentangle the evolutionary state(s) of this group. 

\section{Conclusion}

By combining asteroseismic and spectroscopic measurements for a sample of giant stars we discovered a correlation between A(Li) and $\Delta\Pi_{1}$, whereby super-Li-rich stars are almost universally young RC stars, and Li-normal stars are predominately older RC stars. 

The simplest explanation for this is that (i) there is a Li-enrichment phase before the start of the RC, either near the RGB tip or during the core flashing phase, and (ii) Li is depleted during the early phases of RC evolution. The exact time of the Li depletion, which could be in the very early phases of RC evolution ($\sim 0.3\%$ of RC lifetime) or later during the early RC (first $\sim 40\%$ of RC evolution) is indistinguishable with our current data. If it occurs in the very early RC phase then it implies that \emph{all} low-mass stars go through a SLR phase. If it occurs later then $\sim1\%$ go through a SLR phase. More data is required to separate these scenarios. Further, the relation between A(Li) and $\Delta\Pi_{1}$ will help to trace the transition of giants from the tip of RGB to RC phase where stars burn helium at the center quiescently.

\section{Acknowledgement}

We thank anonymous referees for their comments which has improved manuscript. We gratefully acknowledge the entire team of Kepler space telescope which is  funded by NASA's Science Mission.  This work has made use of data from the European Space Agency (ESA) mission {\it Gaia} (\url{https://www.cosmos.esa.int/gaia}) .
We have also used publicly available data from LAMOST funded by Chinese Academy of Science. S.W.C. acknowledges federal funding from the Australian Research Council through Future Fellowship FT160100046 and Discovery Project DP190102431. This research was supported by use of the Nectar Research Cloud, a collaborative Australian research platform supported by the National Collaborative Research Infrastructure Strategy (NCRIS). We thank Evgenii Neumerzhitckii for the use of his plotting routine. MV acknowledge support from NASA grant 80NSSC18K1582.Y.B.K acknowledge the support from NSFC grant 11850410437.

\appendix
\section{Possible biases in stellar sample}
\label{appx:bias}

Of our 59 RC giants 30 are  taken from \cite{Tajitsu2017}.This forms our low-Li sample of stars as their A(Li) are based on high resolution spectra. The \cite{Tajitsu2017} sample, which only used brightness as a selection criterion, was itself taken from the original sample of \cite{Mosser2012cat} who derived $\Delta\Pi_{1}$ for 95 RC stars that had long cadence observations of the Kepler field. 

At no stage did we select on the basis of $\Delta\Pi_{1}$. Our A(Li) values are however limited in a couple of cases. Firstly, the abundances from the LAMOST medium-resolution spectra are limited to A(Li)~$\gtrsim 0.7$~dex (Sec.~\ref{sec:mrsabunds}). This does not affect the detection of Li-rich and SLR stars, since their abundances are above this limit, but it biases against low-Li stars. Secondly, we used the sample from \cite{Singh2019l} whose Li measurements were restricted only to very strong Li lines due to low spectral resolution (LAMOST, R~$\sim1800$). This resulted in a bias towards SLR stars in their sample, which is mapped to our sample. By combining these samples biased to Li-rich stars with the low-Li sample of \cite{Tajitsu2017}, we cover the whole range of A(Li).

Super-Li-rich stars are central to our main result (Sec.~\ref{sec:results} and \ref{sec:discussion}). The Kepler field is known to have 26 SLR stars in total (\citealt{Singh2019l}; also Table~\ref{tab:tab1}), they are rare objects. Of these, we have 15 in our sample with $\Delta\Pi_{1}$ measurements. As such our sample represents a very substantial fraction ($58\%$) of all the known SLR stars in the Kepler field. 

A possible bias against observing low $\Delta\Pi_{1}$ values was reported by \cite{Constantino2015}, based on comparisons between models and the \cite{Mosser2012c} sample. At low $\Delta\Pi_{1}$ their models predict more stars than are observed. This could be a problem with the models or the observations, or both. As mentioned, our sample contains 58\% of all known Kepler field SLR stars. We have $\Delta\Pi_{1}$ for all of these stars, and our main result (Sec.~\ref{sec:results}) is that they are strongly peaked at low values. This shows that we, and others (see Sec.~\ref{sec:sample} for $\Delta\Pi_{1}$ sources), have been able to measure low values of $\Delta\Pi_{1}$ in the majority of these stars. We see no reason why the detectability of $\Delta\Pi_{1}$ should depend on Li abundance.

\begin{figure}[h]
\centering
\includegraphics[width=0.6\textwidth]{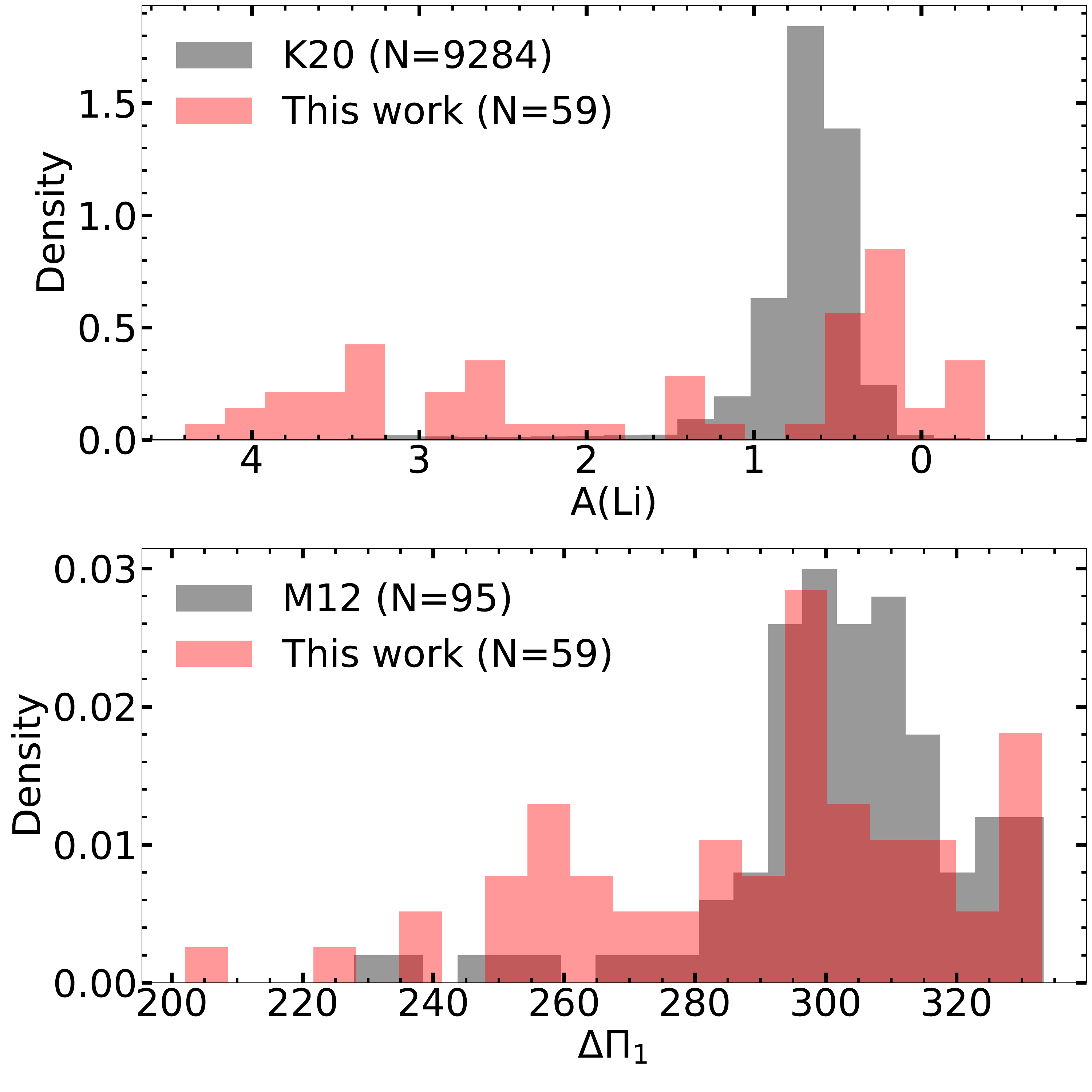}
\caption{Histogram showing the distribution of A(Li) (top) and $\Delta \Pi_{1}$ (bottom) of our RC sample compared to the large Galah sample (K20; \citealt{Bharatn2020}) and the \cite{Mosser2012cat} (M12) catalogue. We note that at low A(Li) the K20 sample has a systematic shift relative to the \cite{Tajitsu2017} sample, of $\sim +0.3$~dex. This could be due to the lower resolution and SNR of GALAH. The bias to high Li can be seen in our sample, as can the unusual peak in $\Delta \Pi_{1}$ at $\sim 255$~s associated with the SLR stars (our main result).}
\label{fig:hists}
\end{figure}

In Figure~\ref{fig:hists} we provide histograms comparing the distributions of $\Delta\Pi_{1}$ and A(Li) in our sample with large survey samples. Our sample covers the whole range of A(Li) and $\Delta\Pi_{1}$ present in the larger samples. As discussed above, our sample has proportionally more (S)LR stars than an unbiased survey, which is seen clearly in Fig.~\ref{fig:hists}. The $\Delta\Pi_{1}$ distribution, despite being independent of our sampling, has an unusual peak at low $\Delta\Pi_{1} \sim 255$~s -- this is our main result, that SLR stars have predominately low $\Delta\Pi_{1}$.

\end{document}